\documentclass[aip,jcp,reprint,floatfix,citeautoscript]{revtex4-2}

\newcommand\temptitle{
A streamlined molecular-dynamics workflow for computing solubilities of molecular and ionic crystals}

\usepackage{geometry}
\usepackage{cmap}
\usepackage[T1]{fontenc}
\usepackage{times}
\usepackage{helvet}
\usepackage[UKenglish]{babel}
\usepackage{csquotes}
\usepackage{newtxtext}
\PassOptionsToPackage{protrusion=true,tracking,kerning,final}{microtype}
\usepackage[slantedGreek]{newtxmath}
\usepackage{microtype}

\DeclareSymbolFont{UPM}{U}{eur}{m}{n}
\DeclareMathSymbol{\uppartial}{0}{UPM}{"40}

\usepackage{graphicx}%

\usepackage[dvipsnames, table]{xcolor}

\usepackage[version=4,arrows=pgf]{mhchem}
\usepackage[modules=symbols]{chemmacros}

\usepackage{booktabs}

\usepackage{comment}

\usepackage{xpatch}
\makeatletter
\xpatchcmd{\@ssect@ltx}{\@xsect}{\protected@edef\@currentlabelname{#8}\@xsect}{}{}%
\xpatchcmd{\@sect@ltx}{\@xsect}{\protected@edef\@currentlabelname{#8}\@xsect}{}{}%
\makeatother

\usepackage[pdftex, bookmarks, pdffitwindow=false, pdfstartview=FitH, pdfdisplaydoctitle, colorlinks, plainpages=false, pdftitle={\temptitle},pdfauthor={}, pdfpagelabels, hypertexnames, citecolor={blue!50!black},linkcolor={blue!50!black}, urlcolor={blue!50!black}, pdflang={en}, hyperfootnotes=false, breaklinks]{hyperref}

\usepackage{siunitx}
\DeclareSIUnit\angstrom{\text{\AA}}
\DeclareSIUnit\calorie{\text{cal}}
\DeclareSIQualifier\NaCl{NaCl}
\DeclareSIQualifier\water{water}

\usepackage{setspace}
\usepackage[labelsep=space,justification=RaggedRight,labelfont={bf},font={footnotesize,stretch=1.15}]{caption}

\usepackage{amsmath}
\usepackage{flafter}

\usepackage{mleftright}

\usepackage[nodayofweek]{datetime}
\newdateformat{myDate}{\THEDAY\ \monthname[\THEMONTH] \THEYEAR}

\def\avg#1{\ensuremath{\mleft\langle #1 \mright\rangle}}

\newcommand{\der}{\ensuremath{\mathrm{d}}}

\newcommand{\kb}{k_{\rm B}}

\newcommand{\bk}{\ensuremath{\textit{\textbf{k}}}}

\newcommand{\figLabel}[1]{\textsf{(\MakeLowercase{#1})}}  
\newcommand{\figLabelCapt}[1]{(\MakeLowercase{{#1}})}
\newcommand{\refSub}[2]{\hyperref[#2]{\ref{#2}\figLabelCapt{#1}}}
\newcommand{\figref}[1]{Fig.~\ref{#1}}
\newcommand{\figrefsub}[2]{Fig.~\refSub{#2}{#1}}

\newcommand{\subfigimg}[3][,]{%
  \setbox1=\hbox{\includegraphics[#1]{#3}}%
  \leavevmode\rlap{\usebox1}%
  \rlap{\hspace*{0pt}\raisebox{\dimexpr\ht1-0.75\baselineskip}{#2}}%
  \phantom{\usebox1}%
}

\begin{document}

\title{\temptitle} 

\author{Aleks Reinhardt}
\email{ar732@cam.ac.uk}
\affiliation{Yusuf Hamied Department of Chemistry, University of Cambridge, Lensfield Road, Cambridge, CB2 1EW, United Kingdom}

\author{Pin Yu Chew}
\affiliation{Yusuf Hamied Department of Chemistry, University of Cambridge, Lensfield Road, Cambridge, CB2 1EW, United Kingdom}

\author{Bingqing Cheng}
\email{bingqing.cheng@ist.ac.at}
\affiliation{Institute of Science and Technology Austria, Am Campus 1, 3400 Klosterneuburg, Austria}

\date{\myDate\today}

\raggedbottom

\begin{abstract}
Computing the solubility of crystals in a solvent using atomistic simulations is notoriously challenging due to the complexities and convergence issues associated with free-energy methods, as well as the slow equilibration in direct-coexistence simulations. This paper introduces a molecular-dynamics workflow that simplifies and robustly computes the solubility of molecular or ionic crystals. This method is considerably more straightforward than the state-of-the-art, as we have streamlined and optimised each step of the process. Specifically, we calculate the chemical potential of the crystal using the gas-phase molecule as a reference state, and employ the S0 method to determine the concentration dependence of the chemical potential of the solute. We use this workflow to predict the solubilities of sodium chloride in water, urea polymorphs in water, and paracetamol polymorphs in both water and ethanol. Our findings indicate that the predicted solubility is sensitive to the chosen potential energy surface. Furthermore, we note that the harmonic approximation often fails for both molecular crystals and gas molecules at or above room temperature, and that the assumption of an ideal solution becomes less valid for highly soluble substances.
\end{abstract}

\maketitle

\raggedbottom

\section{Introduction}

The solubility quantifies the maximum amount of a material, known as the solute, that can be dissolved in a solvent at equilibrium.
A knowledge of the solubility is crucial for a wide range of applications, ranging from pharmaceutical drug formulation to crystal growth, and from chemical synthesis to the phase separation of mixtures~\cite{Aguiar1967, Gerard1999, Cheney2010}.
For example, to ensure that a drug is absorbed at a suitable rate, it should have an appropriately high or low solubility~\cite{Byrn1995},
so it is useful to estimate the solubility as part of the screening process even before the drug is synthesised experimentally.
A range of approaches have been adopted in the literature to predict solubilities~\cite{Skyner2015}, from informatics-based methods~\cite{Ye2021b} to explicit-solvent computational models~\cite{Li2017,Li2018,Khanna2020,Khanna2023,Bellucci2019,Sanz2007,Paluch2010, Moucka2011, Lisal2005, Paluch2015, Espinosa2016, Benavides2016, Boothroyd2018,Dybeck2023}.

However, predicting solubility computationally poses challenges for several reasons. 
First, the thermodynamic equilibrium between the solute and the solution is governed by both energetic and entropic factors. 
Addressing the entropic components often demands intricate and costly computations. 
Second, it is not straightforward to design the thermodynamic cycles necessary to perform these computations.
Finally, it is important to model the molecular interactions accurately. 
Empirical force fields are often parameterised to describe either the liquid or the solid phase of a material accurately, but capturing both simultaneously is much more challenging.  %

The solubility of a crystal corresponds to the solution concentration $c$ at which there is an equilibrium between the solid solute and the solution; at a known pressure $P$ and temperature $T$, the equilibrium is reached when the chemical potentials of the two phases are equal, $\mu^\text{crystal}(P,\,T)=\mu^\text{sol}(P,\,T,\,c)$.
In principle, it may be possible to determine this equilibrium using direct-coexistence simulations~\cite{Opitz1974, Ladd1977} by placing the crystal in the same simulation box as a solution, and determining what the concentration of the solution is once the amount of crystal no longer changes~\cite{Joung2008, RamirezGarcia2023}.
However, such simulations are difficult in practice, since crystal dissolution and precipitation are complex kinetic processes which usually require equilibration times beyond typical timescales of molecular-dynamics (MD) simulations, particularly so when either solutes or solvents are large molecules and the systems of necessity entail large numbers of atoms.
Moreover, many atomic and molecular systems have numerous polymorphs, i.e.~several distinct possible crystal structures.
Although typical solubility ratios of different crystalline polymorphs are within a factor of two, there are exceptions~\cite{Pudipeddi2005}, and each polymorph can in principle have significantly different solubility properties~\cite{Aguiar1967,Cheney2010}. 
Predicting which polymorph will be the most stable or the most soluble under specific conditions requires a consideration of numerous possible crystal structures, further complicating  computational efforts.

To avoid direct-coexistence simulations, one can independently compute the free energies of the crystal~\cite{Frenkel1984,Cheng2018computing,Aragones2012} 
and the solution~\cite{Li2017, Kirkwood1951,Cheng2022,Dawass2019}.
For atomic crystals, the Einstein crystal method~\cite{Frenkel1984,Vega2008} can be used to obtain the absolute free energy by exploiting artificial thermodynamic integration~\cite{Frenkel1984} to switch between the Einstein crystal potential ($U_0$), for which each atom is attached to its equilibrium position by a harmonic spring, and the potential of interest ($U_1$).
For molecular crystals of flexible molecules, the extended Einstein crystal method was developed~\cite{Li2017,Li2018,Gobbo2019,Bellucci2019},
where the molecular crystal is transformed into an Einstein crystal of independent molecules via a multi-step workflow, including turning on additional harmonic restraints on selected atoms of each molecule to control the orientation of the molecules, subsequently turning on Van der Waals and electrostatic interactions between molecules, and finally turning off all harmonic restraints.
This workflow can be somewhat convoluted, and it can be difficult to determine which atoms needs to be tethered and what spring strength is needed on each tethered atom.

The free energy of the molecule in the solution can be computed as the free-energy difference between a solvated and a gas-phase molecule.
In a dilute solution, this can be determined by computing the chemical potential of a single solute molecule in a box of solvent using artificial  thermodynamic integration by switching on the interactions between the solute and the solvent~\cite{Li2017}. 
In practice, such an approach suffers from origin singularities when the particle is first inserted~\cite{Simonson1993}.
This singularity can be circumvented using the cavity method: a cavity can be switched on and increased in size to accommodate the solute molecule, the solute molecule then added, and finally the cavity is gradually switched off~\cite{Li2017}.
For sufficiently dilute solutions, an ideal-dilute solution approximation can be used to estimate the chemical potential at other concentrations; in other words, if Henry's law applies to the solute, $\mu^\text{sol}(x)=\mu^\text{sol}(x\to0)+\kb T \ln x$, where $x$ is the mole fraction of the solute. 
If the solute is not very poorly soluble, it is likely that deviations from ideality will necessitate several laborious calculations of the solvation free energy as the concentration increases~\cite{Bellucci2019}.

The chemical potentials of the solute in solution and in the crystal must have a common origin. In the procedure outlined above, this is achieved by the stepwise calculation of the extended Einstein crystal's free energy. Since the calculation passes through a freely rotating state, the rotational partition function cancels out, and as long as the intramolecular partition functions in the crystal and the solution are the same, a consistent thermodynamic state is obtained~\cite{Li2017, Bellucci2019}.
In the context of computing solid--liquid phase equilibria, it has been suggested that a simpler alternative approach can be used which avoids this effective stepwise solid--fluid transformation~\cite{Khanna2020,Khanna2023}; by computing the absolute free energy of the molecule in the gas phase explicitly, the liquid or solution free energy can be determined relative to the same origin as the molecular crystal.

In this work, we introduce a set of steps that simplify the calculation of solubilities.
We use a combination of thermodynamic integration (TI)~\cite{Frenkel1984}, free-energy perturbation (FEP)~\cite{Zwanzig1954} and the S0 method~\cite{Cheng2022} to compute the chemical potentials required.
In our approach, we employ a Debye crystal reference for solid phases, reducing the need for manual intervention compared to the extended Einstein-crystal method previously discussed. We show how we can compute chemical potentials with a common baseline by using a gas-phase Debye molecule reference. We showcase this workflow by calculating the solubilities of an ionic solid (sodium chloride) and molecular crystals with varied solubilities, including paracetamol in water, paracetamol in ethanol, and urea in water.

\begin{figure*}
    \centering
   \includegraphics[]{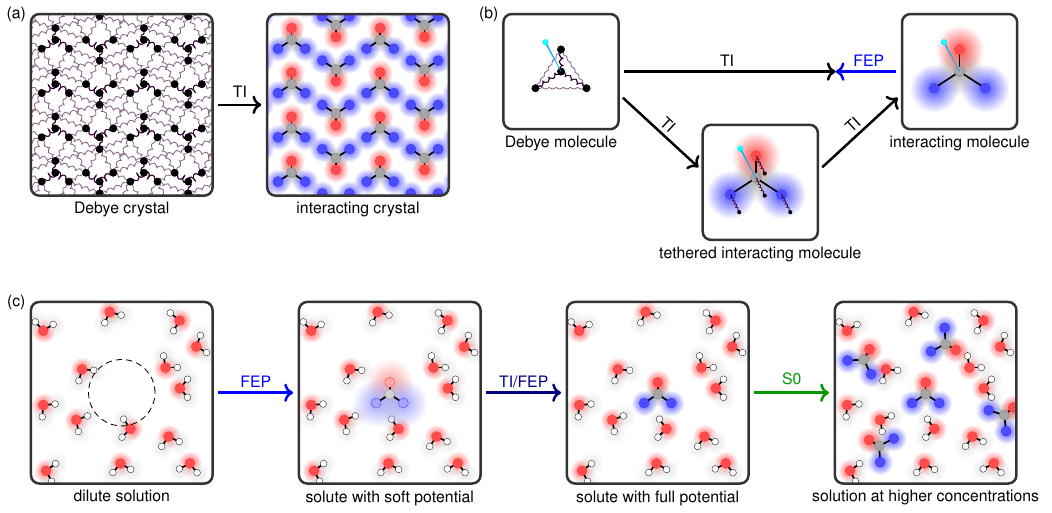}
    \caption{A schematic illustration of the molecular-dynamics workflow for determining chemical potentials. 
    \figLabelCapt{a} Crystalline solute. We use the normal-mode approximation to construct a reference Debye crystal with a known free energy, and then use thermodynamic integration (TI) to switch off the Debye harmonic interaction as the potential of interest is switched on.
    \figLabelCapt{b} Isolated molecule. We construct a normal-mode Debye-molecule landscape with a known free energy. We constrain the molecule's centre of mass and rotational motion, shown pictorially with a cyan pin, and account for these contributions analytically. We then follow one of two routes: we either (1) use artificial thermodynamic integration until most of the potential is switched on and compute the final part by a free-energy perturbation (FEP) method, or (2) use thermodynamic integration to a tethered interacting molecule, and then another thermodynamic integration to switch off the tethering.
    \figLabelCapt{c} Solution. We start with the pure solvent or a very dilute solution. We use a free-energy perturbation to add an additional solute molecule with a soft potential, and then use either thermodynamic integration or further free-energy perturbations to switch the soft potential to the fully interacting system. Finally, we use the S0 method to compute how the chemical potential changes from this initial low concentration as the solute is concentrated.
    }
    \label{fig:method}
\end{figure*}

\section{Methods}\label{sec:methods}

The general workflow is illustrated in \figref{fig:method}.
The absolute chemical potentials of the crystalline phase, the ideal-gas molecule and the solute are each computed separately.
By `absolute', we mean that the chemical potentials have a common origin, irrespective of the reference state from which they are computed.
Specifically, the baseline of the chemical potential for a system comprising molecules with internal bonding, irrespective of the phase, is perhaps most naturally taken to be the chemical potential of an isolated molecule with intramolecular interactions. 
For instance, when determining the chemical potential of a crystalline phase relative to the isolated atoms, we should deduct the baseline chemical potential of the isolated molecule.
This procedure mirrors constructing a thermodynamic cycle between the crystal and the solution via the gas-phase molecule state.  

For the crystal, we use a reference Debye harmonic crystal [Subsec.~\ref{meth:crystal}].
We then determine the baseline chemical potential of an isolated molecule in the gas phase; to do this, we again use a harmonic reference state in which all pairs of atoms are connected with harmonic springs, and separately add analytical chemical potentials of translational and rotational degrees of freedom [Subsec.~\ref{meth:gas-mol}].
Finally, for the solute, we split the calculation into two parts. 
We first compute the solvation free energy at a certain low concentration using a combination of TI and FEP [\S~\ref{meth:solvation}], and then determine the chemical potential of the solution as a function of concentration with the S0 method [\S~\ref{meth:S0}].

\subsection{Crystalline states}\label{meth:crystal}

We use the Debye crystal reference~\cite{Gao2000,Habershon2011b,Cheng2018computing,Chew2023phasediags} as the starting point of TI.
This reference gives a chemical potential baseline of isolated atoms; we account for this in the next step of the calculation by shifting the baseline to that of isolated molecules.
The Debye crystal has all pairs of atoms connected with harmonic springs (as schematically shown in  \figrefsub{fig:method}{a}), and it has the same mass-weighted hessian matrix as the physical crystal.
The hessian can be computed by expanding the potential energy surface around the (local) minimum-energy state in a numerical normal-mode approximation~\cite{Cornwell1986}.
The eigenvalues of the hessian matrix, $\kappa_i$, correspond to the square roots of the normal-mode angular frequencies, $\omega_i=\sqrt{\kappa_i}$, where $i$ identifies each of the normal modes.
The Helmholtz energy of the Debye crystal with a constrained centre of mass (CM) at $T_0$ is
\begin{equation}
A_\text{h} =  \kb T_0 \sum_{i=1}^{3N-3}\ln \frac{\hbar \omega_i}{k_\text{B}T_0},
\end{equation}
where the sum excludes the three normal modes with zero eigenvalues corresponding to translations of the entire system.
For TI, we define a simple linear scaling of the overall potential of a system, $U(\lambda) = \lambda U_1+(1-\lambda)U_0$~\cite{Kirkwood1935}, so that the potential changes from the reference harmonic potential $U_0$ to the potential of interest $U_1$ as the parameter $\lambda$ changes from 0 to 1.
Using artificial thermodynamic integration (`$\lambda$-TI'), the Helmholtz energy of the system governed by $U_1$ is~\cite{Frenkel1984,Vega2008,Cheng2018computing, Chew2023phasediags}
\begin{equation}
A_1=A_0 + \int_0^1 \avg{U_1-U_0}_{\lambda}\,\der \lambda.\label{eq:hamTI}
\end{equation}
Once we have the Helmholtz energy, we can obtain the Gibbs energy~\cite{Cheng2018computing}, e.g.~via $G=A+PV$, and hence the chemical potential $\mu=G/N$.  

Since the harmonic approximation is just the potential energy expanded to quadratic terms, at sufficiently low temperatures, the integral is smooth and the procedure is efficient.
As the temperature increases, however, anharmonic effects begin to play a role and the $\lambda$-TI becomes progressively more difficult.
In particular, rotations about single bonds within each molecule can be activated at higher temperatures, and although in principle this should not lead to a discontinuity in $\langle U_1-U_0\rangle_\lambda$ at $\lambda = 1$, the potential energy difference can become large and the numerical integration may require many data points and long equilibration times.
It is thus preferable to perform the $\lambda$-TI at sufficiently low temperatures when the integral is well-behaved.
Once the Gibbs energy is determined at the initial low temperature $T_0$, we can find how it changes with temperature by numerically integrating the Gibbs--Helmholtz equation,
\begin{equation}
 \frac{G(T)}{T} = \frac{G(T_0)}{T_0} -\int_{T_0}^{T} \frac{H(T')}{(T')^2} \,\der T',
 \label{eq-ti-GH}
\end{equation}
where $H=U+PV$ is the enthalpy of the system.

\subsection{Gas-phase molecule}\label{meth:gas-mol}

To obtain a consistent origin for the absolute chemical potentials of the crystal and the solution, we subtract the free energy of the gas-phase molecule from the free energy of the crystal referenced to the isolated atom state.
To compute the free energy of the isolated molecule, we perform a $\lambda$-TI between a harmonic reference and the physical system [\figrefsub{fig:method}{b}].
It is advantageous, and indeed often necessary, to constrain the centre of mass (CM) and the rotational degrees of freedom of the molecule when performing TI, and then one can add explicit free-energy contributions from the free ideal-gas particle to correct for the CM constraint ($A_\text{cm}$), as well as the rigid-body rotations of the entire molecule ($A_\text{rot}$).

The Helmholtz energy associated with the free CM in a volume $V$ at temperature $T$ is
\begin{equation}
     A_\text{cm} =- \kb T \ln\mleft[V \left(\dfrac{m \kb T}{2\uppi\hbar^2}\right)^{3/2}\mright],
    \label{eq:A_cm}
\end{equation}
and the rotational Helmholtz energy is~\cite{McQuarrie2000}
\begin{equation}
     A_\text{rot} = - \kb T \ln\mleft[ \sqrt{\uppi}\left( \frac{2\kb T}{\hbar^2}\right)^{3/2} \sqrt{I_1 I_2 I_3} \mright],
    \label{eq:A_rot}
\end{equation}
where $I_1$, $I_2$ and $I_3$ are the principal components of the moment of inertia tensor.
The rotational partition function sometimes entails a division by a symmetry number to account for the number of equivalent rotations of the molecule;
however, in this framework, we do not do so because the same degeneracies are also ignored when sampling the free energy of molecular crystals with restricted molecular orientations.

When performing the $\lambda$-TI from the reference system to the potential of interest, although in principle one can use any harmonic or ideal-gas reference, choosing a reference harmonic molecule that has the same frequency modes and equilibrium configuration as the real molecule is statistically efficient.
Selecting such a reference system follows the same procedure as setting up the Debye crystal in the previous section, except that a total of 6 degrees of freedom are removed (3 translational and 3 rotational). 
The classical Helmholtz energy of such a constrained Debye molecule at temperature $T_0$ is
\begin{equation}
    A_\text{h} (V,\,T_0) = \kb T_0 \sum_{i=1}^{3N-6} \ln\dfrac{\hbar \omega_i}{\kb T_0},\label{eq:ahar}
\end{equation}
where $\omega_i$ is the angular frequency of normal mode $i$.

Finally, the Helmholtz energy of the real molecule with a fixed CM and orientation can be obtained using $\lambda$-TI (Eq.~\ref{eq:hamTI}).
As with the Debye crystal, in order to improve the numerical convergence of this thermodynamic integration, this step should be performed at some low temperature $T_0$ at which the system is quasi-harmonic.
Although for fairly rigid molecules this procedure works well, for molecules with different conformers or rotating bonds, the integrand can still adopt very large values as $\lambda \to 1$.
To circumvent this, one can envisage two approaches, as illustrated in  \figrefsub{fig:method}{b}.
The first approach is to perform a FEP in lieu of the final stage of the TI from $1-\varepsilon$ to 1, i.e.
\begin{equation}
    A(\lambda=1) - A(\lambda=1-\varepsilon) =
    \kb T_0 \ln\avg{
    \exp\mleft[-\dfrac{\varepsilon(U_0-U_1)}{\kb T_0}\mright]}_{\lambda=1},
    \label{eq:gcut}
\end{equation}
where $\varepsilon$ is a small number close to zero.
The second approach is to perform a TI between the harmonic reference and a tethered molecule (the physical system plus harmonic springs connecting each atom to its equilibrium position), and another TI between the tethered and the untethered molecule.
Finally, the temperature dependence of the free energy from $T_0$ to $T$ can be computed using the Helmholtz-energy analogue of Eq.~\eqref{eq-ti-GH}.

\subsection{Solute in solution}

\subsubsection{Solvation free energy}\label{meth:solvation}
The solvation free energy is the free-energy difference between a solute molecule that is fully interacting with a dilute solution and the same molecule in the ideal-gas state.
To compute it, in principle we can use $\lambda$-TI [Eq.~\eqref{eq:hamTI}] between these two states directly.
Here the potential of interest $U_1$ corresponds to the molecule fully interacting with its surroundings at some concentration of the solution. 
The reference potential $U_0$ corresponds to the solute molecule not interacting with its surroundings, but where the intramolecular energy of the molecule, as well as the interactions between solvents, are the same as in $U_1$. 
Typically, we use a solution with one molecule dissolved in a box of the solvent for this purpose.

Many procedures for the hamiltonian switching are available~\cite{Beutler1994,Moucka2011}.
The simplest is the linear $\lambda$ scaling and get the free energy difference using Eq.~\eqref{eq:hamTI}.
However, a na\"ive implementation of the $\lambda$-TI results in a divergent integrand in Eq.~\eqref{eq:hamTI} as $\lambda\to 0$: when the `ghost' molecule and the rest of the system are not interacting, atoms can overlap and cause infinity in $U_1-U_0$.
To circumvent this problem, one can instead perform a FEP at the end point~\cite{Schmid2023} (as highlighted using a blue arrow in \figrefsub{fig:method}{c}.), i.e.
\begin{equation}
A(\lambda=\varepsilon) - A(\lambda=0) =  - \kb T\ln\avg{\exp\mleft[-\frac{\varepsilon(U_1-U_0)}{\kb T}\mright]}_{\lambda=0},
\end{equation}
and then compute the integral in Eq.~\eqref{eq:hamTI} from $\varepsilon$ to 1 in $\lambda$ only.
 
Another approach that avoids numerical convergence issues in TI is the use of multiple-step FEP using a $\lambda$-dependent soft-core potential.
For example, a possible soft-core functional form of the Lennard-Jones potential is~\cite{Beutler1994}
\begin{equation}
\begin{split}
    U_\text{LJ}^\text{soft} &= \lambda 4 \varepsilon_{\mathrm{LJ}} \Biggl\{
 \left[ \alpha_{\mathrm{LJ}} (1-\lambda)^2 +
\left( \displaystyle\frac{r}{\sigma_{\mathrm{LJ}}} \right)^6 \right]^{-2} \\ & \qquad\qquad\qquad{} -
 \left[ \alpha_{\mathrm{LJ}} (1-\lambda)^2 +
\left( \displaystyle\frac{r}{\sigma_{\mathrm{LJ}}} \right)^6\right]^{-1} \Biggr\},
\end{split}
\label{eq:lj-soft}
\end{equation}
where $\varepsilon_{\mathrm{LJ}}$ and $\sigma_{\mathrm{LJ}}$ are appropriate energy and distance units, and $\alpha_{\mathrm{LJ}}$ controls the switching protocol.
Similarly, a soft-core Coulomb interaction is
\begin{equation}
    U_{\mathrm{C}}^\text{soft} = \lambda \frac{ q_i q_j}{4\uppi \varepsilon_0 \varepsilon_{\mathrm{C}} \left[ \alpha_{\mathrm{C}}
(1-\lambda)^2 + r^2 \right]^{1/2}},
\label{eq:coul-soft}
\end{equation}
where $q_i$ and $q_j$ are the charges on the two interacting atoms, $\varepsilon_0$ is the electric constant, $\varepsilon_{\mathrm{C}}$ is the dielectric constant, and $\alpha_{\mathrm{C}}$ controls the switching.   %

When using multiple-step FEP, one uses a number of intermediate states, for example,
$U(\lambda_i)$ with $\lambda_i = i/M$ for $i \in \{ 0,\,1,\,\ldots,\,M-1\}$.
We have found it to be statistically efficient to combine a forward and a backward FEP at half steps.
For example, to compute the free-energy difference between two potential-energy surfaces $U_{\text{a}\vphantom{b}}$ and $U_\text{b}$, one runs the simulation using the potential $U_{1/2} = (U_{\text{a}\vphantom{b}} + U_\text{b})/2$,
and the estimator is
\begin{equation}
\begin{split}
        A_\text{b} - A_\text{a} & =  
        - \kb T\ln\avg{\exp\mleft[-\dfrac{U_\text{b}-U_{\text{a}\vphantom{b}}}{2\kb T}\mright]}_{U_{1/2}} \\
    & \qquad\qquad{} + \kb T\ln\avg{
    \exp\mleft[-\dfrac{U_{\text{a}\vphantom{b}}-U_\text{b}}{2 \kb T}\mright]}_{U_{1/2}}  .
\end{split}
\label{eq:fep-mid}
\end{equation}

\subsubsection{Concentration-dependent chemical potential}\label{meth:S0}

In the previous step, we computed the solvation free energy, and hence chemical potential $\mu^\text{sol}_0$, at some low concentration $c_0$.
To find how this quantity changes as a function of solution concentration, we use the recently proposed S0 method~\cite{Cheng2022}, as shown schematically by the green arrow in \figrefsub{fig:method}{c}.
The S0 method is based on the thermodynamic relationship between fluctuations in particle numbers and derivatives of the chemical potentials with respect to the molar concentration~\cite{Kirkwood1951}, and only uses the static structure factors computed from equilibrium MD simulations~\cite{Cheng2022}.
Specifically, we perform multiple equilibrium MD simulations at different concentrations and find $\mu^\text{sol}(c)$ by numerical integration with respect to $\ln c$,
\begin{equation}
\begin{split}
    \mu^\text{sol}(c) & = \mu^\text{sol}(c_0) + \kb T \ln(c/c_0) \\
& \qquad   {} + \kb T \int_{\ln c_0 }^{\ln c} \der \ln(c) \left[ \dfrac{1} {  S_{\text{M--M}}^0-S_{\text{M--S}}^0\sqrt{c/c_\text{S}}  }-1    \right],
    \label{eq:S-integral}
\end{split}
\end{equation}
where the subscripts M and S denote solute and solvent molecules, respectively. $S_{\text{M--M}}^0$ is the static structure factor in the $k \to 0$ limit between solute molecules and $S_{\text{M--S}}^0$ is the equivalent between solute and solvent molecules.  

\section{Results}

\subsection{Sodium chloride in water}
\begin{figure}
  \centering
  \begin{tabular}{@{}p{0.45\textwidth}@{\quad}p{0.45\textwidth}@{}}

    \subfigimg[width=\linewidth]{\figLabel{a}}{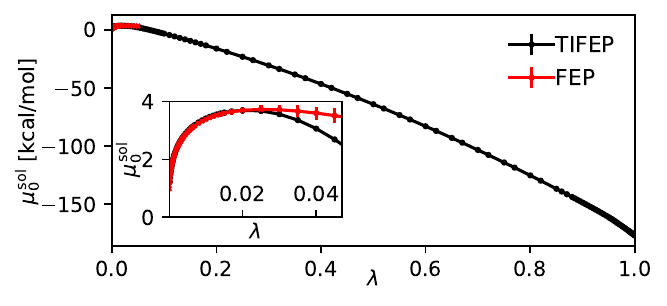}

    \subfigimg[width=\linewidth]{\figLabel{b}}{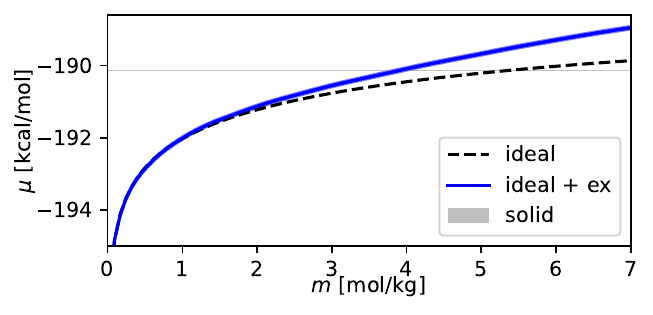}
  \end{tabular}
    \caption{
    \figLabelCapt{a} Chemical potential of solvation $\mu^\text{sol}_0$ of an NaCl ion pair computed using free-energy perturbation (FEP) and a combination of thermodynamic integration and free-energy perturbation (TIFEP). 
    \figLabelCapt{b} The blue curve shows the chemical potentials $\mu^\text{sol}$ for NaCl ion pairs as a function of molality $m$ at \qty{298.15}{\kelvin}.
    The dashed black line is the ideal chemical potential.
    The grey horizontal line indicates the chemical potential of the ion pairs in the solid phase.
    The statistical uncertainties are indicated by the widths of the curves.}
    \label{fig:nacl}
\end{figure}

To model NaCl and water, we used the JC/SPC/E~\cite{Joung2008} force field, with the Lennard-Jones interactions truncated at \qty{1}{\nano\metre} with tail corrections, and long-range Coulomb interactions were treated using a particle--particle particle-mesh solver.
Since NaCl is made up of individual ions, we do not need to explicitly compute the free energy of the gas-phase molecule in this case, as the reference state for the solvation free energy and the crystal free energy are both a pair of ideal-gas NaCl ions with an analytic free energy.

For the solid phase, we performed a $\lambda$-TI from the Debye crystal to the physical system at \qty{298.15}{\kelvin}.
The result is consistent with a $\lambda$-TI first performed at a lower $T$ of \qty{50}{\kelvin} and then an integration along an isobar [Eq.~\ref{eq-ti-GH}].
The chemical potential for each ion pair in the solid phase \qty{298.15}{\kelvin} is \qty{-190.145(15)}{\kilo\calorie\per\mole}. 

To compute the solvation free energy, we used i-PI~\cite{Kapil2019} to perform TI between two states: (i) a pair of Na and Cl ions that are non-interacting and a box of water (with 510 water molecules), and (ii) Na and Cl ions that are fully interacting with water and with each other and a box of water.
Crucially, a combination of a stochastic velocity rescaling thermostat~\cite{Bussi2007} and a weak local Langevin thermostat was used to ensure sufficient equilibration of the ghost molecule and efficient temperature control.
The time step was \qty{1}{\femto\second}, and each MD run lasted for $\sim$\qty{1.5}{\nano\second}.
We used FEP from $\lambda=0$ to $\lambda=0.0001$, and linear $\lambda$ scaling in the TI over a dense grid for the rest.
The FEP at $\lambda=0$ is necessary because the TI integrand is divergent at that point due to the singularity of the LJ and the Coulomb potential related to the ions.
In \figrefsub{fig:nacl}{a}, we show the comparison of the excess chemical potential $\mu^\text{sol}_0(\lambda)$ computed using FEP and the combination of TI and FEP (TIFEP)~\cite{Schmid2023}, as a function of the switching parameter $\lambda$.
At $\lambda < 0.02$, both methods yield similar results.
At higher $\lambda$, however, the direct FEP has a much higher statistical error and a systematic bias.
Our estimated solvation free energy is $\mu^\text{sol}_0 = \qty{-177.30(5)}{\kilo\calorie\per\mole}$.

The final step is to obtain the $\mu^\text{sol}$ of NaCl ion pairs at different concentrations.
Excess chemical potentials for NaCl ion pairs ($\mu^\text{ex}_\text{NaCl}$) were computed using the S0 method (Eq.~\eqref{eq:S-integral}) in Ref.~\onlinecite{Cheng2022}.
Simulations of NaCl water solutions at different molar concentrations were performed using LAMMPS~\cite{Plimpton1995}
at \qty{298.15}{\kelvin} and \qty{1}{\bar}; more details can be found in Ref.~\onlinecite{Cheng2022}.
The ideal chemical potentials is given by the first two terms on the right-hand side of Eq.~\eqref{eq:S-integral}, and is shown by the dashed black curve in \figrefsub{fig:nacl}{b}.
By combining the ideal and excess parts of the chemical potential, we computed $\mu^\text{sol}$ of NaCl ion pairs at different salt molalities $m$ (i.e.~the chemical amount of the solute per unit mass of the solvent), shown as the blue curve in \figrefsub{fig:nacl}{b}.
In the same figure, we also show the chemical potential of the ion pairs in the crystal, $\mu^\text{crystal}$ [grey horizontal line].
The point at which the blue curve and grey line cross gives the solubility of NaCl in water; for this model, we estimate the solubility to be  \qty{3.87(14)}{\mole\NaCl\per\kilo\gram\water}. %
This estimate is lower than the experimental result of \qty{6.15}{\mole\per\kilo\gram} at \qty{25}{\celsius}~\cite{Pinho2005}, but is consistent with previous simulation results computed using osmotic ensemble Monte Carlo~\cite{Moucka2013}, the Bennett acceptance ratio method~\cite{Mester2015activitycoeffs} and thermodynamic integration~\cite{Benavides2016}, all employing the same force field.

\subsection{Urea in water}\label{subsec:urea}

The properties of urea--water mixtures have been investigated using a range of empirical potentials~\cite{Kokubo2007,Kokubo2007b,Mountain2004,Bertran2002,Smith2004}.
For our investigation, we used the CHARMM36~\cite{Klauda2010} parameterisation from CHARMM-GUI~\cite{Jo2008,Lee2016b}, since unlike several other commonly used empirical potentials, we verified that the experimentally known crystalline polymorphs are at least mechanically stable at suitable thermodynamic conditions.
The potential entails combinations of Lennard-Jones and Coulomb interactions, coupled with bond, angle, dihedral and improper constraints that maintain a relatively rigid molecular structure~\cite{MacKerell1998}.
We do not use any further rigid-body constraints. 
We remark that since harmonic bonds are used, simulations should not be run with the Nos\'e--Hoover thermostat~\cite{Hoover1985, Aragones2013}, and alternatives such as stochastic velocity rescaling~\cite{Bussi2007} or the Langevin thermostat should be used instead.
For the solution phase, water molecules are modelled with a potential compatible with the urea CHARMM force field; the water model entails a three-site potential with Coulomb and Lennard-Jones interactions at each water atom, and harmonic bond and angle constraints.

\begin{figure}[!ht]
  \centering
  \begin{tabular}{@{}p{0.45\textwidth}@{\quad}p{0.45\textwidth}@{}}
    \subfigimg[]{}{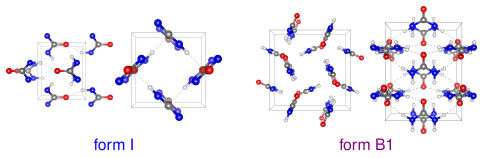}
    \subfigimg[width=\linewidth]{\figLabel{a}}{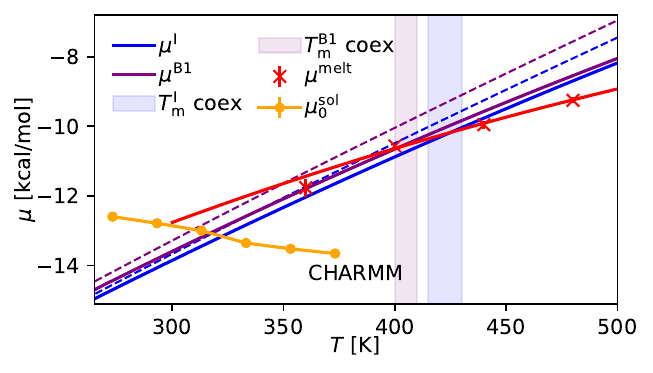}
    \subfigimg[width=\linewidth]{\figLabel{b}}{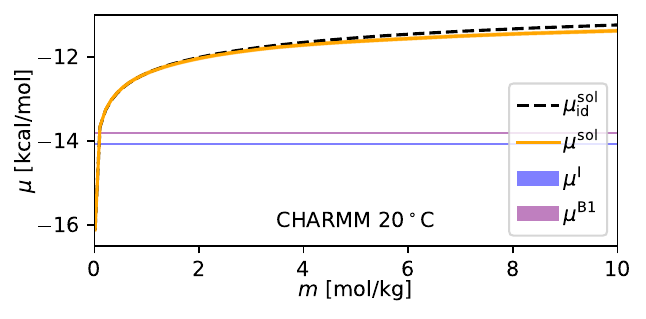}
    \subfigimg[width=\linewidth]{\figLabel{c}}{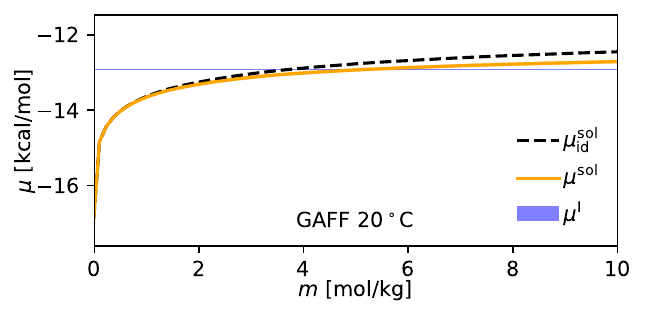}
  \end{tabular}
    \caption{
    \figLabelCapt{a} Chemical potentials of urea in different phases computed using the CHARMM force field.
    The blue and purple solid curves show the chemical potentials of
    crystal form I ($\mu^\mathrm{I}$) and crystal form B1 ($\mu^\mathrm{B1}$), respectively. The dashed curves show the harmonic free energies of the corresponding phases.
    The red crosses with error bars show the estimate of the chemical potential of melted urea ($\mu^\mathrm{melt}$), and the red curve shows the values from the thermodynamic integration of the liquid along the \qty{1}{\bar} isobar.
    The blue and the purple shaded areas show estimates of the melting temperatures of forms I and B1 with uncertainty, computed using the coexistence method.
    The orange symbols show the solvation free energy $\mu^\mathrm{sol}_0$ at the concentration $c_0 = \qty{0.49}{\mole\per\cubic\deci\metre}$ with error estimates.
    \figLabelCapt{b} Chemical potentials of urea in solution of different molalities $m$ (i.e.~the amount of urea dissolved in a kilogram of water) computed using the CHARMM force field.
    \figLabelCapt{c} Chemical potentials of urea in solution of different molalities computed using the GAFF force field.
    In \figLabelCapt{b} and \figLabelCapt{c}, the orange curve shows the chemical potentials $\mu^\mathrm{sol}$ for urea as a function of the urea molality.
    The dashed black line is the ideal chemical potential.
    The blue and purple horizontal lines indicates the chemical potential of phase I and B1.
    The statistical uncertainties are indicated by the width of the curves.
    }
    \label{fig:urea}
\end{figure}

For the bulk of our work, we considered two crystalline polymorphs of urea, namely forms I~\cite{Guth1980} and B1~\cite{Piaggi2018} (\figref{fig:urea}). 
For both polymorphs, we performed a $\lambda$-TI first at a low temperature (\qty{25}{\kelvin}) using a Debye crystal reference, followed by TI in $T$ along the \qty{1}{\bar} isobar.
The free energy computed for the crystals is thus referenced to isolated atoms.
For the gas phase, the workflow for computing the free-energy reference to the isolated atoms is similar to that of crystals, except that both translational and rotation are constrained in the TI simulations. The analytic translational and rotational free-energy corrections (Eq.~\eqref{eq:A_cm} and Eq.~\eqref{eq:A_rot}) are added at the end.
The volume of the gas used in Eq.~\eqref{eq:A_cm} was $1000a_0^3$, where $a_0$ is the Bohr radius.
The chemical potentials of the crystalline polymorphs ($\mu^\mathrm{I}$ and $\mu^\mathrm{B1}$) are calculated by subtracting the free energy of the gas-phase molecule from the free energies of the crystal polymorphs.
These chemical potentials are shown in \figrefsub{fig:urea}{a}.
Consistently with previous work~\cite{Piaggi2018,Olejniczak2009}, form I is the most stable form at ambient pressure.
By way of comparison, the harmonic-approximation chemical potentials $\mu^\mathrm{I}_\text{h}$ and $\mu^\mathrm{B1}_\text{h}$ are also plotted in \figrefsub{fig:urea}{a}; the difference to the true chemical potentials accounting for anharmonicity is rather large in both cases.

The solvation free energy is computed using the multiple-step FEP with the soft-core potentials,
with the switching parameters $\alpha_\mathrm{LJ} = 0.5$ and $\alpha_\mathrm{C} = 10$ in Eqs~\eqref{eq:lj-soft} and~\eqref{eq:coul-soft}.
We used a total of 20 windows, and performed the FEP in the middle step as described in Eq.~\eqref{eq:fep-mid}.
We confirmed that using more windows did not change the estimate.
By contrast, a forward or backward FEP would require many more windows to reach convergence.
Each independent run lasts for $\sim$\qty{5}{\nano\second}.
We used a time step of \qty{1}{\femto\second} since the resulting $\mu^\mathrm{sol}_0$ is consistent with that computed using a smaller time step of \qty{0.1}{\femto\second}. 
The calculations were performed using i-PI~\cite{Kapil2019}.
A combination of a stochastic velocity rescaling thermostat~\cite{Bussi2007} and a weak local Langevin thermostat was used for ergodic sampling of the solute molecule and efficiency.  
The pressure was kept at \qty{1}{\bar}.
As discussed above, the computed solvation free energy is already referenced to the gas-molecule state.

As a sanity check, we computed the solvation free energy of urea in its own melt, which is just the chemical potential of the melt $\mu^\mathrm{melt}$ (the red symbols in \figrefsub{fig:urea}{a}).
$\mu^\mathrm{melt}$ computed at different temperatures agree well with the values obtained from TI of the melted phase with respect to $T$ (Eq.~\eqref{eq-ti-GH}), as shown using the red curve in \figrefsub{fig:urea}{a}.
The crossover between $\mu^\mathrm{melt}$ and $\mu^\mathrm{I}$ or $\mu^\mathrm{B1}$ is the melting point of the specific crystalline phase,
and we estimate $T_\text{m}^\mathrm{I} = \qty{425(7)}{\kelvin}$ and $T_\text{m}^\mathrm{B1} = \qty{405(5)}{\kelvin}$.
We then performed independent direct-coexistence simulations in elongated boxes (3840 atoms for form I with typical box dimensions of \qtyproduct{71 x 22 x 22}{\angstrom}  and 5760 atoms for form B1 with typical box dimensions of \qtyproduct{90 x 27 x 21}{\angstrom}). 
We first determined the equilibrium lattice parameters as a function of temperature at \qty{1}{\bar}, then melted half the box along $x$ at a high temperature whilst keeping the remaining half the molecules frozen, and finally evolving the system at fixed $P_x$ until one phase grew at the expense of the other.
The melting point for form I determined with this method is in the range \qtyrange{415}{430}{\kelvin}, and for form B1 it is in the range \qtyrange{400}{410}{\kelvin}.
Since both ranges are consistent with estimates from free-energy calculations, this suggests that the computed chemical potentials had a consistent choice of baseline, and the correctness of the workflow.
The experimental melting point of form I urea is \qty{406}{\kelvin}~\cite{Miller1934}; the close agreement with simulation is perhaps somewhat surprising given the simplicity of the force field.

\begin{table}
\centering
\begin{tabular}{S[table-format=3]@{\hspace{4mm}}S[table-format=3.2]@{\hspace{5mm}}S[table-format=2.3(3)]@{\hspace{5mm}}S[table-format=2.3(3)]}
\toprule
& & {solubility of}  & {solubility of}\\
{$T$ / \unit{\celsius}} & {$T$ / \unit{\kelvin}} & {I / \unit{\mole\per\kilo\gram}} & {B1 / \unit{\mole\per\kilo\gram}} \\
\midrule
20 & 293.15 & 0.054(2)   & 0.083(3) \\  
40 & 313.15 & 0.236(9)   & 0.36(1)  \\
60 & 333.15 & 1.17(4)  & 1.82(7)  \\
80 & 353.15 & 4.2(1)  & 6.9(2)  \\
100 & 373.15 & 17.2(7) & 33(1)  \\
\bottomrule
\end{tabular}
\caption{Solubilities of urea in water expressed as molalities at different temperatures for the CHARMM potential. 
}
\label{tab:urea-sol}
\end{table}

We computed the solvation free energy of urea in water ($\mu^\mathrm{sol}_0$) at \qty{20}{\celsius} intervals between \qty{0}{\celsius} and \qty{100}{\celsius}.
Once a `ghost' urea molecule was added, the simulation box contained 1 urea molecule and 123 water molecules, corresponding to a concentration of about \qty{0.49}{\mole\per\cubic\deci\metre}.
These results are indicated using orange symbols with error bars in \figrefsub{fig:urea}{a}.
To obtain chemical potentials of the solute at other concentrations, we used the S0 method.
We simulated urea--water mixtures at different molar concentrations using LAMMPS~\cite{Plimpton1995} at the temperatures specified above and a pressure of \qty{1}{\bar}.
The simulation box contained about 10,000 molecules.
We used a time step of \qty{1}{\femto\second}, with runs of $\sim$\qty{1}{\nano\second} each.
To compute $S(\bk)$, we collected a snapshot per 1000 steps of the trajectory.
We used the positions of oxygen atoms as the positions of water molecules,
and the positions of carbon atoms as the positions of urea molecules.
The $S(\bk)$ values were then fitted to the Ornstein--Zernike form~\cite{Cheng2022} with a maximum cutoff in the wave vector $k_\text{cut}^2 = \qty{0.004}{\per\angstrom\squared}$.
The $S^0$ values between urea--urea and urea--water can then be used to compute the concentration-dependent $\mu^\mathrm{sol}(c)$ using Eq.~\eqref{eq:S-integral}.
As an example, $\mu^\mathrm{sol}$ at \qty{20}{\celsius} at different molalities is shown as the orange curve in \figrefsub{fig:urea}{b}.
The shaded area indicates the statistical uncertainty, which principally comes from the error in the estimation of $\mu^\mathrm{sol}_0$.
The ideal chemical potential is shown as the dashed black curve.
Finally, the chemical potentials of the crystalline forms I and B1 are plotted as blue and purple horizontal lines.
As the concentration increases, the solution becomes less ideal.
The crossover points between the orange curve and the two horizontal lines indicate the solubilities of the corresponding phases.
In this case, the ideal-solution assumption turns out to be rather accurate in the solubility prediction.
However, the agreement becomes worse at other temperatures considered; e.g.~at \qty{80}{\celsius}, it would lead to an underestimate in the solubility by about \qty{15}{\percent}.
We summarise the solubilities at other temperatures in Table~\ref{tab:urea-sol}.
The solubilities increase steeply with temperature, and the solubility of form B1 is higher than that of form I at all temperatures considered.

It has been shown that using different urea models can result in considerably different thermodynamic properties in solution~\cite{Kokubo2007b}.
To check the role of the choice of force field in determining the solubility, we used the AMBER-GAFF-ESP-2018 force field (GAFF)~\cite{Ozpinar2010,Spoel2018} for both urea and water.
We performed the solubility calculations for urea form I in water with GAFF, following the same workflow as with CHARMM.\footnote{For the GAFF potential, form B1 readily converts to form B2 at the conditions of interest, and so we have not considered it further.}
In \figrefsub{fig:urea}{c}, we show $\mu^\mathrm{sol}$ at \qty{20}{\celsius} computed using GAFF as a function of urea molality.
At this temperature, the solubility of form I of urea is \qty{5.1(3)}{\mole\per\kilo\gram}, about 100 times the solubility predicted for the CHARMM potential under the same conditions.

Experimentally, urea dissolves very readily in water; the solubility in \qty{100}{\gram} of water ranges between $\sim$\qty{70}{\gram} (\qty{12}{\mole\per\kilo\gram}) at \qty{0}{\celsius} to $\sim$\qty{700}{\gram} (\qty{120}{\mole\per\kilo\gram}) at \qty{100}{\celsius}~\cite{Pinck1925,Miller1934,Halonen2016}.
These solubilities are considerably higher than the CHARMM predictions obtained in  Table~\ref{tab:urea-sol}.
The GAFF solubility at \qty{20}{\celsius} is closer to, but still lower than, the experimental value.
On the other hand, the crystalline phase is less well described by the GAFF potential than by CHARMM; %
at temperatures above \qty{20}{\celsius}, form I is not dynamically stable under the GAFF force field, and readily transforms into an analogue of form III in MD simulations.
These results thus demonstrate the sensitivity of the solubility prediction on the assumed potential energy surface, and the challenge of accurately modelling both the crystalline and the solution phases with the same potential.

\subsection{Paracetamol in water and in ethanol}

We used the CHARMM36~\cite{Klauda2010} parameterisation from CHARMM-GUI~\cite{Jo2008,Lee2016b} for paracetamol, water and ethanol, since all common crystalline polymorphs of paracetamol are at least metastable with this potential.

Specifically, we considered two crystalline polymorphs of paracetamol, namely forms I~\cite{Ou2020} and II~\cite{Thomas2011}. 
For both polymorphs, we computed the free energy of the Debye crystal at \qty{25}{\kelvin} and then used $\lambda$-TI to the potential of interest, followed by a TI in $T$ along the \qty{1}{\bar} isobar.
For both polymorphs, it is important to perform the $\lambda$-TI from the Debye crystal at $T\lesssim \qty{40}{\kelvin}$, as the internal rotational modes become activated at higher $T$, causing numerical problems with the integrand of Eq.~\eqref{eq:hamTI} as $\lambda\to 1$.
We used a time step of \qty{0.25}{\femto\second} and, as with the case of urea above, a combination of a stochastic velocity rescaling thermostat~\cite{Bussi2007} and a weak local Langevin thermostat.

\begin{figure}
  \centering
  \begin{tabular}{@{}p{0.45\textwidth}@{\quad}p{0.45\textwidth}@{}}
    \subfigimg[]{}{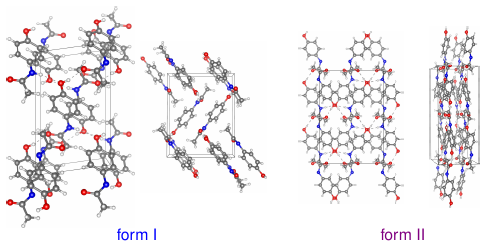}
    \subfigimg[width=\linewidth]{\figLabel{a}}{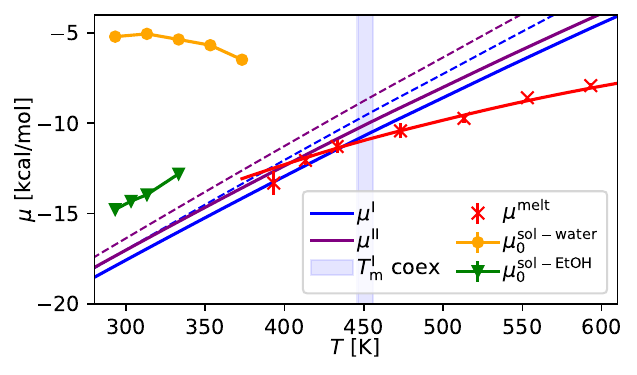}
    \subfigimg[width=\linewidth]{\figLabel{b}}{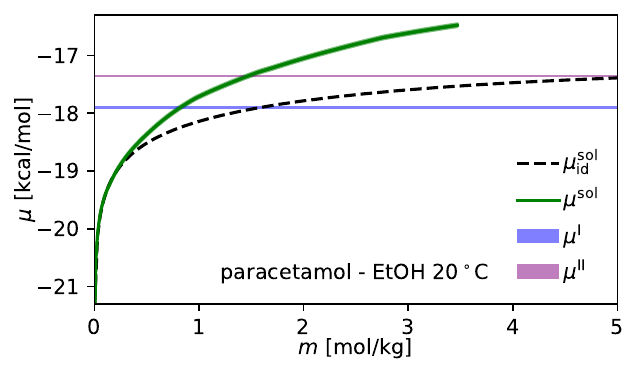}
  \end{tabular}
    \caption{
    \figLabelCapt{a} Chemical potentials of paracetamol in different phases computed using the CHARMM force field.
    The blue and purple solid curves show the chemical potentials of paracetamol in crystal forms I ($\mu^\mathrm{I}$) and II ($\mu^\mathrm{II}$), respectively. The dashed curves show the harmonic free energies of the corresponding phases.
    The red crosses with error bars show the estimate of the chemical potential of the paracetamol melt ($\mu^\mathrm{melt}$), and the red curve shows the values from the TI of the liquid along the \qty{1}{\bar} isobar.
    The blue shaded area shows the estimate of the melting temperature of phase I with statistical uncertainty computed using the interface-pinning method for solid--liquid coexistence.
    The orange symbols show the solvation chemical potentials $\mu^\text{sol--water}_0$ of paracetamol in water at $c_0 = \qty{0.47}{\mole\per\cubic\deci\metre}$ with error estimates,     and the green symbols show the solvation chemical potentials $\mu^\text{sol--EtOH}_0$ in ethanol at $c_0 = \qty{0.23}{\mole\per\cubic\deci\metre}$.
    \figLabelCapt{b} The green curve shows the chemical potentials $\mu^\mathrm{sol}$ for paracetamol dissolved in ethanol (EtOH) as a function of molality $m$ (i.e.~amount of paracetamol per kilogram of ethanol).
    The dashed black line is the ideal chemical potential.
    The blue and purple horizontal lines indicate the chemical potentials of forms I and II, respectively.
    Statistical uncertainties are indicated by the widths of the curves.
    }
    \label{fig:para}
\end{figure}

For the gas phase, we computed the free-energy difference between the Debye molecule reference and the molecule with a constrained CM and rotations at $T=\qty{100}{\kelvin}$.
Because rotations about bonds are possible and lead to different conformers, vanilla $\lambda$-TI is thwarted by divergences in the integrand.
Instead, both the tethering approach and the FEP at the end point, as outlined in the \nameref{sec:methods} section, help to circumvent the issue.
We used a time step of \qty{0.25}{\femto\second},  and the same combined thermostat as for the solid phases.
The analytic translational and rotational free-energy corrections are added in the end.
The volume of the gas used in Eq.~\eqref{eq:A_cm} was $1587a_0^3$. 
The total free energy of the gas molecule at \qty{100}{\kelvin} was estimated to be \qty{-41.84(4)}{\kilo\calorie\per\mole}. 
The free energies at other temperatures were then determined using TI along the \qty{1}{\bar} isobar.

The chemical potentials of the crystalline polymorphs ($\mu^\mathrm{I}$ and $\mu^\mathrm{II}$) are baseline adjusted, so the reference state is the gas-phase molecule.
These chemical potentials are shown in \figrefsub{fig:para}{a}, together with the harmonic-approximated $\mu^\mathrm{I}_\text{h}$ and $\mu^\mathrm{II}_\text{h}$ in dashed lines; the difference is of the order of \qty{1}{\kilo\calorie\per\mole}. 

For computing solvation free energies, we used the multiple-step FEP with the soft-core potentials described in Sec.~\ref{meth:solvation},
with $\alpha_\mathrm{LJ} = 0.5$ and $\alpha_\mathrm{C} = 10$.
We used a total of 20 windows and performed the FEP in the middle step as described above.
We confirmed that using more windows did not change the estimate.
For the solvation free energy of paracetamol in water ($\mu^\text{sol--water}_0$),
the simulation box contained 1 paracetamol molecule and 118 water molecules ($c_0 = \qty{0.47}{\mole\per\cubic\deci\metre}$).
The time step was \qty{1}{\femto\second}, and each independent run lasted about \qty{1.5}{\nano\second}. 
For the solvation free energy of paracetamol in ethanol ($\mu^\text{sol--EtOH}_0$),
we used a simulation box comprising 1 paracetamol molecule and 73 ethanol molecules ($c_0 = \qty{0.23}{\mole\per\cubic\deci\metre}$), and each independent FEP run lasted about \qty{5}{\nano\second}.
These solvation free energies ($\mu^\text{sol--water}_0$ and $\mu^\text{sol--EtOH}_0$) are shown in \figrefsub{fig:para}{a}.
The large gap between the two sets of solvation energies indicates that paracetamol solvation in ethanol is rather more favourable than in water when using the CHARMM force field.

In addition, we computed the solvation free energy of paracetamol in its own melt, $\mu^\mathrm{melt}$, shown using the the red symbols in \figrefsub{fig:para}{a}.
These values agree well with the TI results of the melt (Eq.~\eqref{eq-ti-GH}) [red curve in \figrefsub{fig:para}{a}].
We estimate the melting points of form I and II to be $T_\text{m}^\mathrm{I} = \qty{440(11)}{\kelvin}$ and $T_\text{m}^\mathrm{II} = \qty{407(11)}{\kelvin}$.
As a sanity check on our free-energy calculations, we also computed the melting point of form I of paracetamol using interface pinning~\cite{Pedersen2013,Pedersen2013b,Pedersen2015}.
We took an elongated box of crystal form I (roughly %
\qtyproduct{107x29x35}{\angstrom} with 504 molecules, with $\alpha=\gamma=\ang{90}$ and $\beta\approx\ang{98}$), equilibrated it at \qty{1}{\bar} and a range of temperatures to determine optimal box parameters, melted half the box at a high temperature and locally equilibrated the system.
We then performed interface-pinning simulations in which only the $x$ component of the pressure was coupled to a barostat.
We used a Steinhardt--Ten Wolde-style order parameter~\cite{Steinhardt1983,TenWolde1996} on the carbonyl oxygen to classify molecules as being either in the crystal or the liquid phase, and then, using PLUMED~\cite{Tribello2014}, applied a harmonic restraint with spring constant $\kappa$ to bias the number of crystalline molecules to be close to half the total molecules in the system, $N_0\approx N/2$.
We computed the chemical-potential difference between form I and the melt via $\upDelta \mu = \kappa(N_0-\langle N_\text{cryst} \rangle)$~\cite{Pedersen2013, Chew2023phasediags},
by determining the average number of crystalline molecules as a function of temperature.
Using this approach, we estimate that the melting point, at which $\upDelta \mu=0$, is \qty{451(5)}{\kelvin}.
These results are consistent with \figrefsub{fig:para}{a}, giving us confidence that all the relevant factors have been considered in the free-energy calculations.
The experimental melting point of form I paracetamol is \qty{442}{\kelvin}~\cite{Ledru2007} (\qty{169}{\celsius}), in close agreement with simulation despite the assumptions of the force field.

For the concentration dependence of $\mu^\mathrm{sol}$, we used the S0 method. 
We simulated paracetamol--water and paracetamol--ethanol mixtures at different molar concentrations using LAMMPS~\cite{Plimpton1995},
although for paracetamol--water this proved not to be necessary, since at low concentrations the solution is nearly ideal.
The simulation box contained about 10,000 molecules.
The time step was \qty{1}{\femto\second} and each run lasted about \qty{1}{\nano\second}.
Snapshots where taken every 1000 steps, and the co-ordinates of oxygen atoms in water, nitrogen atoms in paracetamol, and the carbon bonded to oxygen in ethanol were used for recording the positions of the corresponding molecules.
In the estimation of $S^0$, the cutoff in the wave vector was $k_\text{cut}^2 = \qty{0.0025}{\per\angstrom\squared}$.

We show in \figrefsub{fig:para}{b} the solution chemical potential $\mu^\mathrm{sol}$ for paracetamol in ethanol at \qty{20}{\celsius} as a function of the paracetamol molality [green curve].
The chemical potentials of the two crystalline polymorphs are plotted as blue and purple horizontal lines.
At higher concentrations, the solution becomes less ideal, and indeed the ideal-solution assumption would lead to an overestimate in the solubilities by a factor of two to three.
The solubilities of paracetamol in water or ethanol at all temperatures considered are provided in Tables~\ref{tab:para-sol-water} and \ref{tab:para-sol-EtOH}.
The solubilities increase with temperature in both water and ethanol, and the solubility of form II is about double that of form I under all conditions considered.

\begin{table}
\centering
\begin{tabular}{S[table-format=3]@{\hspace{4mm}}S[table-format=3.2]@{\hspace{5mm}}S[table-format=1.1(2)e-1]@{\hspace{5mm}}S[table-format=1.1(2)e-1]}
\toprule
& & {solubility of}  & {solubility of}\\
{$T$ / \unit{\celsius}} &  {$T$ / \unit{\kelvin}} & {I / \unit{\mole\per\kilo\gram}} & {II / \unit{\mole\per\kilo\gram}} \\
\midrule
20 & 293.15 & 3.5(5)e-8   & 9.1(10)e-8 \\  
40 & 313.15 & 5.1(8)e-07 & 1.3(2)e-06  \\
60 & 333.15 & 1.1(1)e-05& 2.5(3)e-05  \\
80 & 353.15 & 1.6(2)e-04  & 3.6(4)e-04  \\
100 & 373.15 & 3.3(4)e-03 & 7.2(8)e-03  \\
\bottomrule
\end{tabular}
\caption{Solubility of paracetamol in water expressed as molalities at different temperatures.}
\label{tab:para-sol-water}
\end{table}

\begin{table}
\centering
\begin{tabular}{S[table-format=2]@{\hspace{4mm}}S[table-format=3.2]@{\hspace{7mm}}S[table-format=1.2(2)]@{\hspace{5mm}}S[table-format=2.2(2)]}
\toprule
 & & {solubility of}  & {solubility of}\\
{$T$ / \unit{\celsius}} & {$T$ / \unit{\kelvin}} & {I / \unit{\mole\per\kilo\gram}} & {II / \unit{\mole\per\kilo\gram}} \\
\midrule
20 & 293.15 & 0.82(3) & 1.47(5) \\
30 & 303.15 & 0.95(4) & 1.79(7) \\
40 & 313.15 & 1.34(5) & 2.7(1) \\
\bottomrule
\end{tabular}
\caption{Solubility of paracetamol in ethanol expressed as molalities at different temperatures.}
\label{tab:para-sol-EtOH}
\end{table}

The CHARMM22 potential with SwissParam~\cite{Zoete2011} parameters has been used to identify the source of stabilisation of different polymorphs of paracetamol relative to DFT-level calculations~\cite{Rossi2016}; this work illustrates that empirical force fields likely do not capture all the relevant physics.
However, the CHARMM36 force field with optimised partial charges has been shown to be a satisfactory potential for studying the nucleation of paracetamol in acetonitrile~\cite{Stojakovic2017,Gobbo2018}, and this reparameterised potential has also been used to investigate the solubility of paracetamol in ethanol~\cite{Bellucci2019} using the cavity method.
In this work, previous experimental values~\cite{Granberg1999,Jimnez2006,Perlovich2006, Mitchell2010} were averaged to obtain a solubility mole fraction of \num{0.0585(40)} at \qty{20}{\celsius}, or a molality of \qty{1.35}{\mole\per\kilo\gram}.
The solubility of paracetamol with the CHARMM36 potential with reoptimised partial charges was \qty{0.085(14)}~\cite{Bellucci2019} (or \qty{2.0}{\mole\per\kilo\gram}), a slight overestimate compared to the experimental result.
By contrast, using the CHARMM-GUI parameterisation of the charges results in a slight underestimate of the solubility ($m=\qty{0.82}{\mole\per\kilo\gram}$, or $x=0.036$). 
Both parameterisations appear to be reasonable, but the solubility is very sensitive to the details of the model's parameters.
However, the solubility of paracetamol in water at \qty{20}{\celsius} is $m=\qty{0.0845}{\mole\per\kilo\gram}$~\cite{Granberg1999} (or $x=0.0015$), increasing to $\sim$$m=\qty{0.48}{\mole\per\kilo\gram}$ ($x=0.0086$) at \qty{70}{\celsius}~\cite{Grant1984}, which is very different from the values we have obtained, suggesting that the interactions of paracetamol with water are accounted for less well.   %

\section{Conclusions}

In summary, we present a workflow that enables simple and robust computation of solubilities of molecular or ionic crystals.
As illustrated in \figref{fig:method}, we compute the free energies of the gas, the crystal and the solution phases separately.
The workflow mainly uses thermodynamic integration from reference systems with known free energies to physical systems, with free-energy perturbations incorporated at the end points of the TI to avoid numerical issues and to increase efficiency.
Compared to the state-of-the-art methods~\cite{Li2017,Li2018,Khanna2023,Gobbo2019,Bellucci2019,Dybeck2023},
we streamline many steps.
For example, we compute the chemical potential of the crystal using the gas-phase molecule as a baseline, which avoids the convoluted many-step switching in the extended Einstein crystal method~\cite{Li2017,Li2018}. 
Using Debye crystals or Debye molecules as the starting points of the TI ensures high statistical efficiency, as these reference potential-energy surfaces closely resemble the potentials of interest.
Constraining the CM and the rotation of the gas-phase molecule further reduces the sampling space.
For the solute, the FEP at half steps (Eq.~\eqref{eq:fep-mid}) requires fewer windows than forward or backward FEP.
We use the S0 method to compute the concentration dependence of the chemical potential of the solute, which overcomes the ideal-solution assumption and avoids separate solvation free-energy calculations at different concentrations that would otherwise be necessary~\cite{Bellucci2019}.

We have applied the workflow to systems of a range of solubilities: sodium chloride in water, urea in water, paracetamol in water, and paracetamol in ethanol.
We additionally computed the solid-liquid equilibria for urea and paracetamol using the same workflow, for validation.
From these calculations, we have observed that the harmonic approximation for both the gas molecules and the crystals often breaks down at higher temperatures. 
Even at room temperature, for molecular crystals, the error can be of the order of a \unit{\kilo\calorie\per\mole}, as in the case of urea and paracetamol. 
The ideal-solution assumption can also fail: it is reasonably accurate for dilute solutions, but can result in an error of a factor of two in the solubility predictions of more soluble substances, such as paracetamol in ethanol.

We have shown that quantitative solubility predictions hinge on the accuracy of the potentials chosen.  %
For instance, the solubility of urea in water at \qty{20}{\celsius} predicted using GAFF is two orders of magnitude higher than the CHARMM analogue.
Solubility calculations could thus provide a useful tool for parameterising empirical potentials.
Beyond enhancing empirical force fields, an alternative approach entails the use of machine-learning potentials (MLPs), which capture the accuracy of quantum-mechanical calculations, but at a considerably reduced computational cost~\cite{Deringer2019}.
Given that MLPs have the capacity to model accurately the intricate interactions at the atomic and molecular levels that are crucial for describing molecular solutions, they stand out as a promising tool for refining solubility predictions. Nonetheless, due to the higher computational demands of MLPs compared to classical force fields, integrating them with an efficient workflow, such as ours, becomes essential to ensure the tractability of calculations.

In this paper, we have focused on classical chemical potentials, but of course one could also account for the influence of nuclear quantum effects (NQEs).
For instance, to obtain the chemical potentials of the molecule in solution, one can still use the workflow presented in \figrefsub{fig:method}{c}
while using the path-integral molecular dynamics (PIMD) formalism~\cite{Marx1996} to represent the whole system.
NQEs on the chemical potentials of the gas or the crystalline phases can be taken into account by integrating the quantum centroid virial kinetic energy with respect to the fictitious mass from the infinite mass to the physical masses using PIMD simulations~\cite{Ceriotti2013,Cheng2016nuclear,Cheng2018hydrogen}.

We expect that our workflow for calculating solubilities will be useful across various technologically significant systems. 
For example, for electrolytes, the ability to compute solubilities easily may deepen our understanding of the solvation of conducting ions in batteries.
Similarly, in drug design, a clearer insight into solubility properties of drugs would enable drug delivery and absorption to be optimised.
Moreover, the method can shed light on the precipitation of crystals from solutions, the behaviour of co-solvents, as well as extraction and purification methods in the chemical industry. 

\bigskip

\begin{acknowledgments}
AR and BC acknowledge resources provided by the Cambridge Tier-2 system operated by the University of Cambridge Research Computing Service funded by EPSRC Tier-2 capital grant EP/P020259/1.  
\end{acknowledgments}

\section*{Data availability statement}
All original data generated for the study are in the
SI repository \url{https://github.com/} (url to be inserted).

%

\end{document}